\documentclass[%
 reprint,
 amsmath,amssymb,
 aps,
]{revtex4-1}

\usepackage{graphicx}% Include figure files
\usepackage{dcolumn}% Align table columns on decimal point
\usepackage{bm}% bold math
\usepackage{subfigure}
\begin{document}

\preprint{APS/123-QED}

\title{Enhancement of spontaneous entanglement generation via coherent quantum feedback}% Force line breaks with \\
\author{Bin Zhang} \email{binzhang220@163.com}
\author{Sujian You}
\author{Mei Lu}
\affiliation{College of Mathematics and Physics, Fujian University of Technology, Fuzhou 350118, China}%Lines break automatically or can be forced with \\

\date{\today}% It is always \today, today,

\begin{abstract}
  We investigate the entanglement dynamics of two two-level emitters (qubits) mediated by a semi-infinite, one-dimensional (1D) photonic
  waveguide. The coupling of each qubit to the waveguide is chiral, which depends on the propagation direction of light. The finite
  end of the waveguide is terminated by a perfect mirror, such that coherent quantum feedback is introduced to the system. We show that the
  chirally generated entanglement between the qubits can be preserved by controlling the time delay of the feedback. Moreover, when the time
  delay is negligible, the qubit-qubit reduced system evolves within the strong-coupling regime and the qubits can be almost maximally entangled.
  We also analyze the robustness of the protocol against variations of some relevant parameters.
\end{abstract}

\pacs{Valid PACS appear here}% PACS, the Physics and Astronomy
\maketitle

%\tableofcontents

\section{\label{sec:level1}INTRODUCTION}

Efficient manipulation of the interaction between quantum emitters and quantized electromagnetic (EM) field is a central goal
in quantum optics. This is of particular importance for quantum computation and quantum information, where many schemes require
delicate control of light-matter interaction. One-dimensional (1D) waveguide QED systems provide an excellent platform to engineer
various light-matter interactions, and can be a promising candidate for quantum information processing \cite{kimble,chang,kolchin,gauthier,lemonde}.
Here, a quantum emitter interacts with light in either of the two propagation directions of the waveguide, with conventionally
the same coupling strength. However, this symmetry is violated after the experimental demonstration of chiral light-matter interfaces
in a range of different waveguide systems \cite{junge,petersen,connor,song,feber,coles}. In these systems, the coupling between emitters
and light depends on the propagation direction of light, which is a manifestation of the so-called optical spin-orbit coupling \cite{bliokh}.
Chiral light-matter interaction has promoted the fundamental research on light and found many interesting applications in quantum information
processing \cite{lodahl,ramos,vermersch,yan,Li,Berman}.

On the other hand, the creation of quantum entanglement has always been a fundamental task for quantum computation and quantum information
processing \cite{horodecki}. In recent years, there has been much attention paid on establishing quantum entanglement in
regular \cite{vidal,cano,gonzalez,ballestero,facchi,zheng,Xin} and chiral \cite{daley,moreno,kien,mirza,schotland,Pzoller,Pguimond}
waveguide QED systems. In particular, it is shown that, by mere coupling to an infinite waveguide, qubits could establish effective
interaction and two-qubit entanglement can be generated spontaneously \cite{vidal,cano,gonzalez,ballestero,facchi,moreno,kien}. Moreover,
compared to the non-chiral case, chiral waveguide setups can increase the maximal achievable entanglement and enhance the robustness of the
generated state \cite{moreno}. Recently, schemes for spontaneous entanglement generation of multiqubit \cite{mirza} and two-photon
\cite{schotland} entanglement have also been proposed.

Inspired by previous studies, we investigate the spontaneous entanglement generation between two two-level emitters (qubits) chirally coupled
to a 1D semi-infinite waveguide \cite{chen,matthew,tufarelli,kim,fang,Pguimond,guimond}. The finite end of the waveguide is terminated
by a perfect mirror, which may reflect photons moving towards it and thus induce time-delayed emitter-photon interaction. Indeed, this system
is a very simple example of coherent quantum feedback that suffices for the occurrence of nonexponential atomic decay \cite{tufarelli} and
non-Markovian dynamics \cite{kim,fang}. In this paper, we first derive the delay derivative equations (DDEs) for the evolution of the two-qubit reduced
system, and study the entanglement dynamics for the qubits. We show that, apart from the non-Markovian entanglement revival phenomena,
the coherent feedback can also be used to preserve the two-qubit entanglement with appropriate choices of time delay and chirality. For the intermediate extent of chirality, quasi-steady two-qubit entangled states can be generated, with the entanglement death time much longer than that in the infinite waveguide situation. Furthermore, when the time delay of the feedback is negligible and the coupling tends to be non-chiral, the two-qubit reduced system evolves in the strong coupling regime, such that the qubits can be almost maximally entangled. We also discuss the robustness of the
generated entanglement against variations of time-delay, position and detuning.

This paper is organized as follows: in Sec. II we address the dynamics of our system. In Sec. III we report our results. In Sec. IV, we discuss
the robustness of our protocol. Finally, in Sec. V we close by summarizing our conclusions. In the Appendix, we give a detailed derivation
of the delay derivative equations for the evolution of the qubits.

\section{\label{sec:level2}THEORETICAL MODEL}

\begin{figure}[t]
\includegraphics[width=7cm]{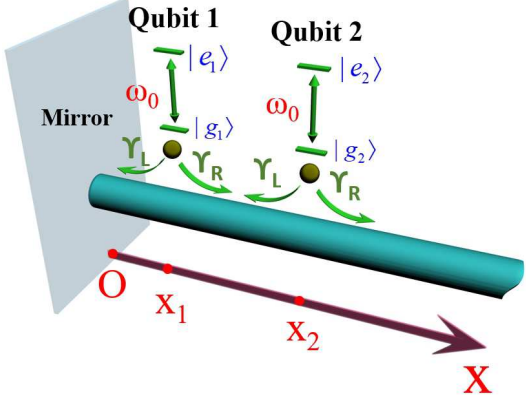}
\caption{\label{setup} Schematic of the setup. Two identical two-level emitters (qubits) are coupled to a semi-infinite waveguide,
which is terminated by a perfect mirror at x = 0. The qubits are placed at $x=x_1$ and $x=x_2$, respectively. Each qubit couples to the left-
(right-) moving guided modes with a rate $\gamma_L$ ($\gamma_R$).}
\end{figure}

We consider two identical two-level emitters 1 and 2 coupled to a 1D semi-infinite waveguide along the x axis, as shown in Fig.~\ref{setup}.
The waveguide is terminated at $x=0$ by a perfect mirror, while the emitters 1 and 2 are placed separately at $x=x_1$ and $x=x_2$.
The ground and excited states of the $m$th ($m=1,2$) emitter are denoted by $|e_m\rangle$ and $|g_m\rangle$, with a transition frequency
$\omega_0$. Both emitters are dipole coupled to a continuum of EM modes of the waveguide, with each mode characterised by wave vector $k$,
frequency $\omega_k$, and bosonic creation (annihilation) operator $a_k^\dag$ ($a_k$). The Hamiltonian for the whole system can be expressed
as ($\hbar=1$)
\begin{equation}
H=H_S+H_B+H_{\mathrm{int}},
\end{equation}
with $H_S=\sum_m\omega_0|e_{m}\rangle\langle e_{m}|$ and $H_B=\int_0^{k_c} \mathrm{d}k \;\omega_k a_k^\dag a_k$ representing the Hamiltonians
for the emitters and the waveguide modes, respectively. Here, $k_c$ stands for the cutoff wave vector of the waveguide. The interaction term
in the rotating wave approximation is given by
\begin{eqnarray}
H_{\mathrm{int}}=&&\sum_{m}\int_{0}^{k_c} \mathrm{d} k\;(g_{ k\scriptscriptstyle L} e^{-ikx_m}+g_{k\scriptscriptstyle R}e^{i(\pi+kx_m)})
\sigma_m^\dag a_k+
\mathrm{H.c.},\nonumber\\
\end{eqnarray}
where $g_{ k\scriptscriptstyle L}$ ($g_{ k\scriptscriptstyle R}$) is the coupling strength of each emitter to the guided mode k that propagates
to the left (right), and $\sigma_m^\dag=|e_{m}\rangle\langle g_{m}|$. Note that there is a phase difference $\Delta \varphi=(\pi+2kx_m)$ between
the two couplings, which is associated with the propagation distance and mirror reflection of the feedback photon. Following the standard
approaches to similar systems \cite{witthaut,shen,fan}, we make the assumption that the photon dispersion relation is approximately linear
around the emitter's transition frequency $\omega_0$, namely $\omega_k\approx \omega_0+v(k-k_0)$, where $v$ is the group velocity of light,
and $\omega_{k_0}=\omega_0$. Moreover, as only a narrow bandwidth of the EM modes in vicinity $\omega_0$ is relevant to the emitter-waveguide
interaction, we can extend the range of k to $(-\infty,+\infty)$. Consequently, the coupling strengths are independent of the frequency.
Following the procedure of Ref. \cite{tufarelli}, we find that $g_{ k\scriptscriptstyle J}=i\sqrt{\gamma_{\scriptscriptstyle J}v/2\pi}$ $(J=L,R)$,
where $\gamma_{L}$ and $\gamma_{R}$ are the spontaneous emission rates that the emitters decay to the left- and right-moving waveguide modes,
respectively. For chiral couplings, we have $\gamma_{L}\neq\gamma_{R}$, and the ratio $\gamma_{R}/\gamma_{L}$ can be employed as a measure of
chirality. Notice that $[H,\sum_m |e_{m}\rangle\langle e_{m}|+\int_0^{k_c} \mathrm{d}k \;a_k^\dag a_k]=0$, the total excitation number of the
system is conserved throughout the evolution.

Suppose that initially only one of the emitters is excited, and the waveguide is in the vacuum state, then the whole system would evolve
in the one-excitation subspace, and the wave function is given by
\begin{eqnarray}\label{3}
|\Psi(t)\rangle=&&\alpha(t)|e_1\rangle |g_2\rangle|0\rangle+\beta(t)|g_1\rangle |e_2\rangle |0\rangle\nonumber\\
&&+|g_1\rangle |g_2\rangle\int \mathrm{d}k \; \phi(k,t)a_k^\dag|0\rangle ,
\end{eqnarray}
where $\phi(k,t)$ is the field amplitude in k space, and $|0\rangle$ is the field vacuum state.
Substituting Eq.~(\ref{3}) into the Schr$\ddot{\mathrm{o}}$dinger equation $i\partial_t|\Psi\rangle=H|\Psi\rangle$, we obtain the
following differential equations:
\begin{eqnarray}\label{4}
\dot{\alpha}(t)=-i\omega_0 \alpha(t)-i\int \mathrm{d}k\; (G_{ k\scriptscriptstyle L1}+G_{ k\scriptscriptstyle R1})\phi(k,t),
\end{eqnarray}
\begin{eqnarray}\label{5}
\dot{\beta}(t)=-i\omega_0 \beta(t)-i\int \mathrm{d}k\; (G_{ k\scriptscriptstyle L2}+G_{ k\scriptscriptstyle R2})\phi(k,t),
\end{eqnarray}
\begin{eqnarray}\label{6}
\partial_t\phi(k,t)&=&-i\omega_k\phi(k,t)-i(G^*_{ k\scriptscriptstyle L1}+G_{ k\scriptscriptstyle R1}^*)\alpha(t)\nonumber\\
&&-i(G^*_{ k\scriptscriptstyle L2}+G_{ k\scriptscriptstyle R2}^*)\beta(t),
\end{eqnarray}
where $G_{ k\scriptscriptstyle Lm}= i\sqrt{\gamma_L v/2\pi}\;e^{-ikx_m}$ and
$G_{ k\scriptscriptstyle Rm}= i\sqrt{\gamma_R v/2\pi}\;e^{i(\pi+kx_m)}$.
By integrating Eq.~(\ref{6}) in terms of $\alpha(t)$ and $\beta(t)$, and substituting it into Eqs.~(\ref{4}) and (\ref{5}), one
observes that
\begin{eqnarray}
\label{7}
\dot{\alpha}(t)=&&-\frac{\gamma_L+\gamma_R}{2}\alpha(t)-\gamma_Le^{ik_0(x_2-x_1)}\beta(t-\frac{x_2-x_1}{v})\nonumber\\
&&\times\theta(t-\frac{x_2-x_1}{v})+\sqrt{\gamma_L\gamma_R}\;e^{ik_02x_1}\alpha(t-\frac{2x_1}{v}) \nonumber\\
&&\times\theta(t-\frac{2x_1}{v})+\sqrt{\gamma_L\gamma_R}\;e^{ik_0(x_1+x_2)}\beta(t-\frac{x_1+x_2}{v})\nonumber\\
&&\times\theta(t-\frac{x_1+x_2}{v}),
\end{eqnarray}
\begin{eqnarray}
\label{8}
\dot{\beta}(t)=&&-\frac{\gamma_L+\gamma_R}{2}\beta(t)-\gamma_Re^{ik_0(x_2-x_1)}\alpha(t-\frac{x_2-x_1}{v})\nonumber\\
&&\times\theta(t-\frac{x_2-x_1}{v})+\sqrt{\gamma_L\gamma_R}\;e^{ik_02x_2}\beta(t-\frac{2x_2}{v}) \nonumber\\
&&\times\theta(t-\frac{2x_2}{v})+\sqrt{\gamma_L\gamma_R}\;e^{ik_0(x_1+x_2)}\alpha(t-\frac{x_1+x_2}{v})\nonumber\\
&&\times\theta(t-\frac{x_1+x_2}{v}),
\end{eqnarray}
where $\theta(t)$ is the Heaviside step function. A detailed derivation of Eqs.~(\ref{7}) and (\ref{8}) is outlined in Appendix.

The above two delay differential equations (DDEs) describe waveguide mediated two-qubit coupling in one excitation subspace.
It is worthwhile to stress that, unlike the master equation formalism, the Markovian approximation is not performed in deriving
the above equations. Therefore, our formalism is more general in the sense that the system can be studied both in the Markovian and
non-Markovian regions. The physical meaning of these equations is clear: The first terms on the right-hand side of Eqs.~(\ref{7}) and (\ref{8})
describe the trivial spontaneous emission of qubits 1 and 2 to the bidirectional EM modes, with a total decay rate $\gamma=\gamma_L+\gamma_R$.
The second terms correspond to the waveguide-mediated effective dipole-dipole coupling between the qubits, which has been predicted in the
infinite waveguide situation \cite{moreno}. Indeed, these two equations are equivalent to the master equation derived in Ref. \cite{moreno},
as long as we omit the last two terms on the right of the equations and set $(x_2-x_1)/v=0$. The latter is the usual Markovian approximation,
in which the retardation effect in the effective qubit-qubit interaction is neglected. The last two terms in Eqs.~(\ref{7}) and (\ref{8})
describe the coherent quantum feedback brought about by the mirror. For instance, the third term in Eq.~(\ref{7}) refers to the process that
a left-moving photon emitted by qubit 1 gets reflected by the mirror and reinteracts with qubit 1, with the delay $2x_1/v$ being the round-trip
propagation time of the photon. The remainder terms can be interpreted analogously. As we shall see later, these terms play a crucial role for
the enhancement of the qubit-qubit entanglement.

\section{\label{sec:level3}NUMERICAL RESULTS AND DISCUSSION}

We proceed by considering the two-qubit entanglement dynamics in our system. In terms of the basis
$\{|e_1\rangle |e_2\rangle,|e_1\rangle |g_2\rangle,$ $|g_1\rangle |e_2\rangle,|g_1\rangle |g_2\rangle\}$,
the reduced density matrix for the qubits is given by
\begin{eqnarray}
\rho^{12}(t) &=& tr_w\{|\Psi(t)\rangle\langle\Psi(t)|\}\nonumber\\
&=&\left( \begin{array}{cccc}
0 & 0 & 0 & 0 \\
0 & |\alpha(t)|^2 & \alpha^*(t)\beta(t) & 0 \\
0 & \alpha(t)\beta^*(t) & |\beta(t)|^2 & 0 \\
0 & 0 & 0 & 1-|\alpha(t)|^2-|\beta(t)|^2
\end{array} \right),\nonumber\\
\end{eqnarray}
where $tr_w$ refers to the trace over the EM modes. In order to characterize the entanglement of the qubits, we calculate the
concurrence \cite{wootters} for $\rho_{12}$, which is given by $C(t)=2|\alpha(t)\beta(t)|$. Notice that concurrence ranges from 0,
which refers to a separable state, to 1, which refers to a maximally entangled state. We also define the quantity $t_d\equiv2x_1/v$
as the time delay of the feedback, and $\tau$ as the time at which maximum cocurrence emerges. For simplicity, hereafter we always
assume that the qubits are initially prepared in the state $|e_1\rangle |g_2\rangle$, and that qubit 1
is placed at the node of the central EM mode (i.e., $k_0x_1=2n\pi$ for some integer $n$).

We first study the influence of coherent feedback on two-qubit entanglement dynamics, where the retardation effect is considered.
In this case, the presence of the mirror induces memory effects of the reservoir, as well as time-delayed information
backflow to the qubits \cite{kim,fang}, so that the system exhibits non-Markovianity. Numerical simulations suggest that, when the
time delay $t_d$ takes intermediate values (with $t_d\geq\tau$), the coherent feedback could not increase the maximal achievable
entanglement. Nevertheless, one can still observe non-Markovian behavior of concurrence.

\begin{figure}[t]
\includegraphics[width=6cm]{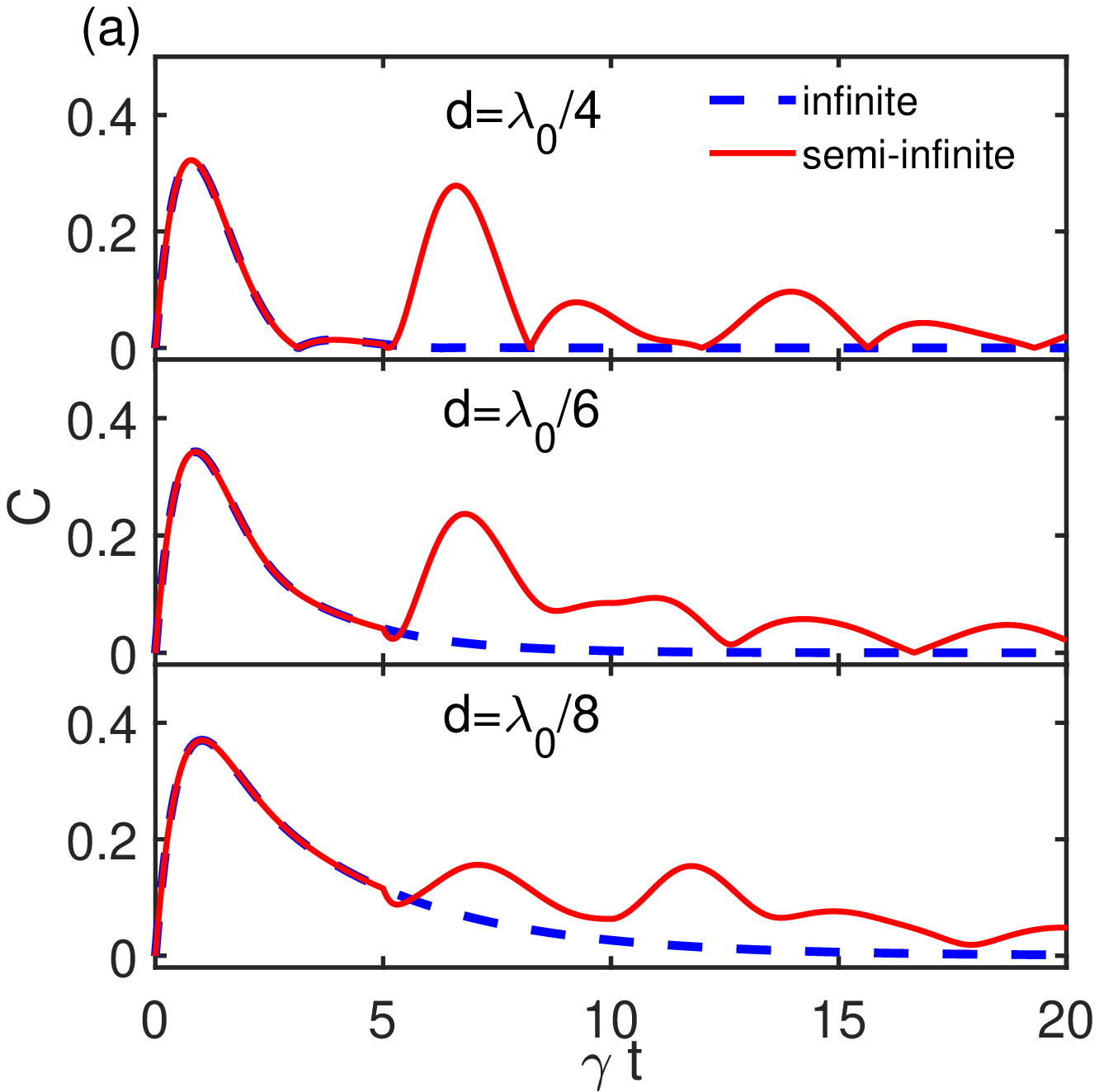}

\includegraphics[width=6cm]{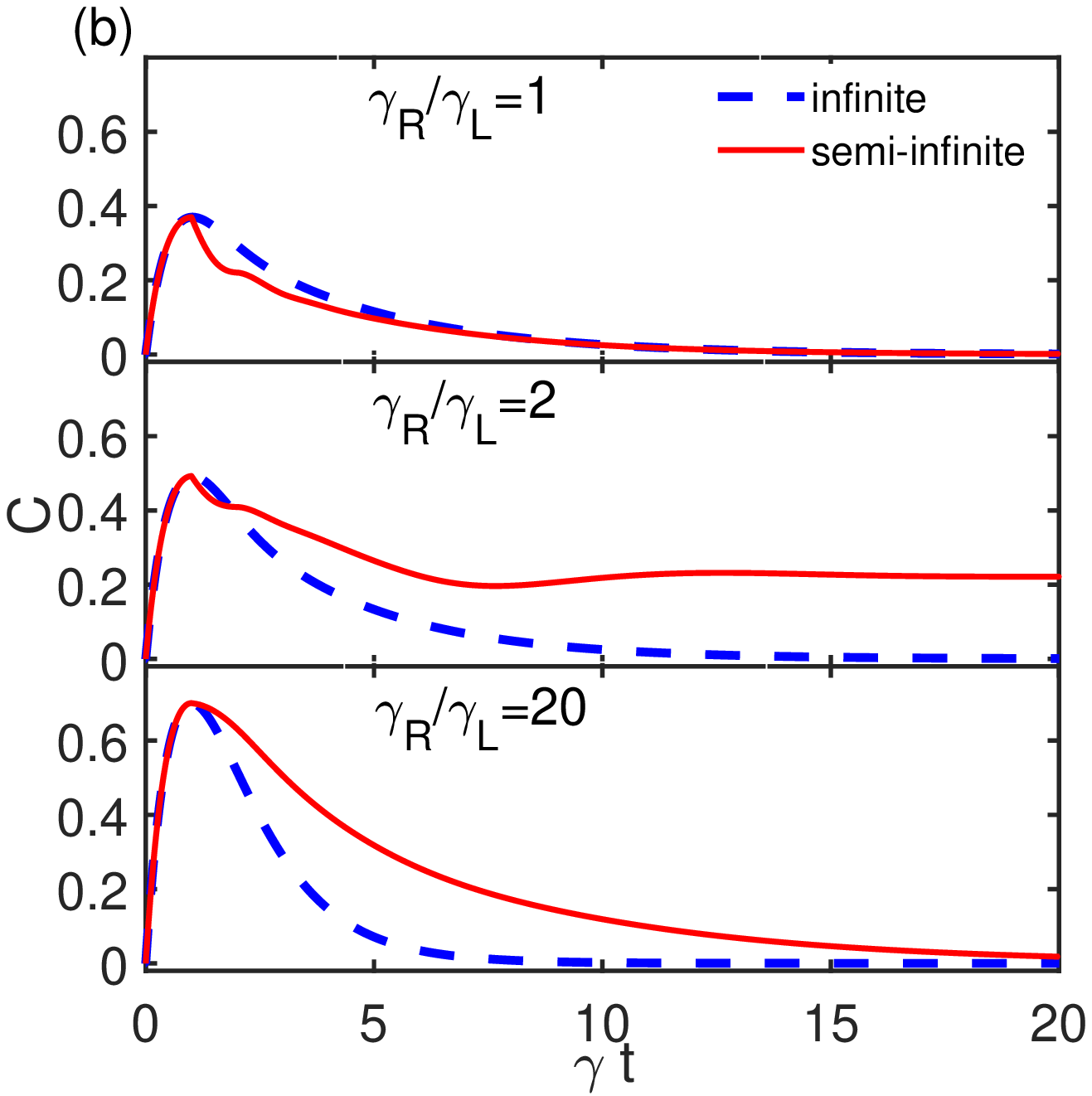}
\caption{\label{chiral1} Concurrence as a function of time t (in units of $\gamma^{-1}$) when the semi-infinite waveguide system evolves
in the non-Markovian regime. For comparison, we also plot the evolution of concurrence in infinite waveguide system under the same conditions.
(a) Entanglement revival phenomena for different qubit-qubit separations, with the coupling being non-chiral. Here we set the time delay
$\gamma t_d=5$. (b) Entanglement preservation by the quantum feedback for different values of $\gamma_R/\gamma_L$, where the time delay
$t_d\approx\tau$ and the qubit-qubit separation $d= \lambda_0/8$. }
\end{figure}

In Fig.~\ref{chiral1}(a), we set $\gamma_L=\gamma_R$ (non-chiral coupling) and plot concurrence $C$ against $\gamma t$ for different
separations between the qubits, with the time delay $t_d=5/\gamma $. Initially, the evolution of concurrence is exactly the
same as that of the infinite waveguide system. However, at times $t\geq t_d$, when the left-moving photon emitted by qubit 1 has completed
the round trip, it can be reabsorbed by the system and entangle the qubits once again. Thereby, unlike the infinite waveguide
system, where concurrence decreases monotonically after rising to maximum value, the semi-infinite setup witnesses the revivals
of entanglement during the disentanglement process. Similar phenomena can also be observed in chiral coupling situations.

\begin{figure*}[t]
\includegraphics[width=6cm]{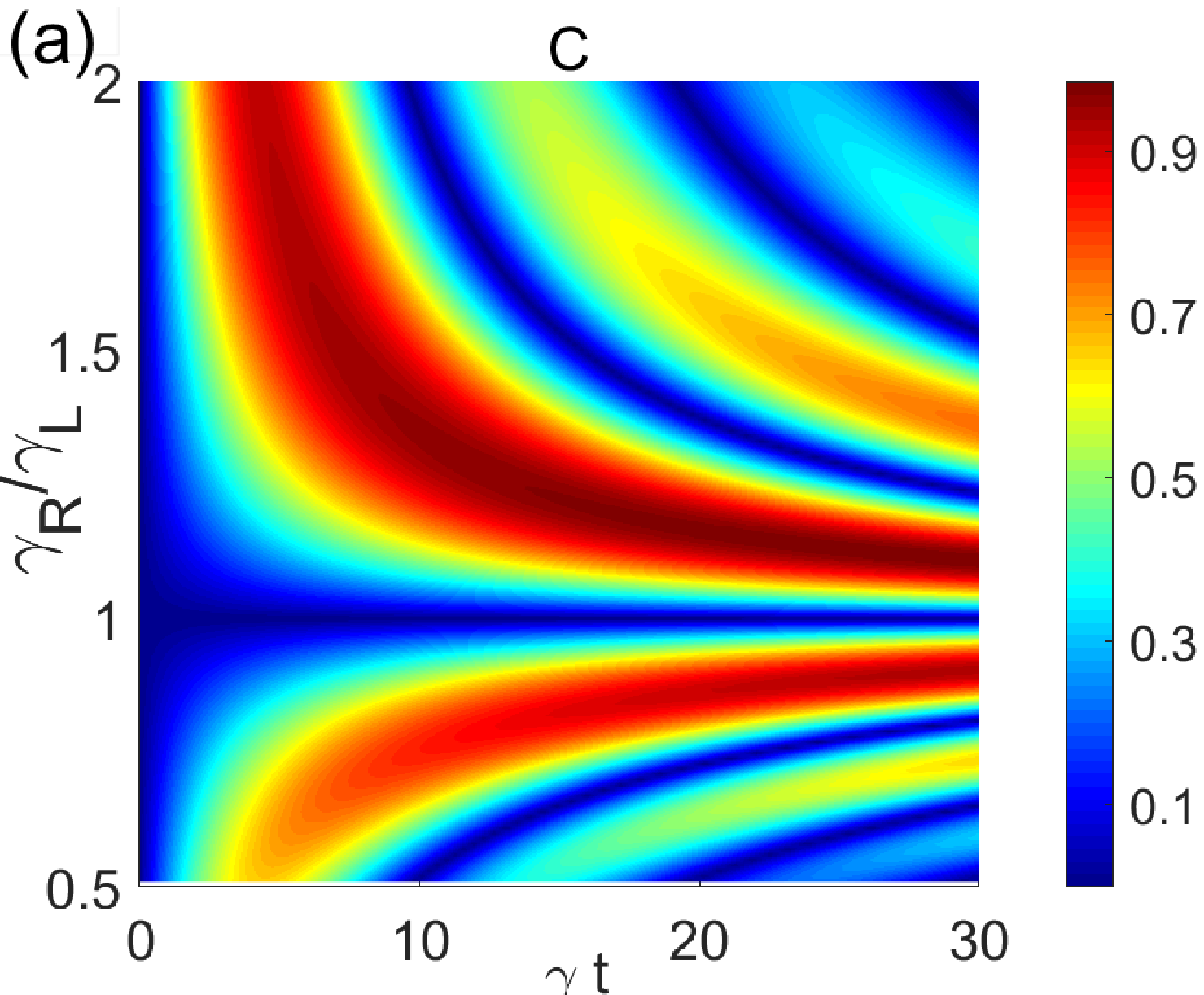}
\includegraphics[width=6cm]{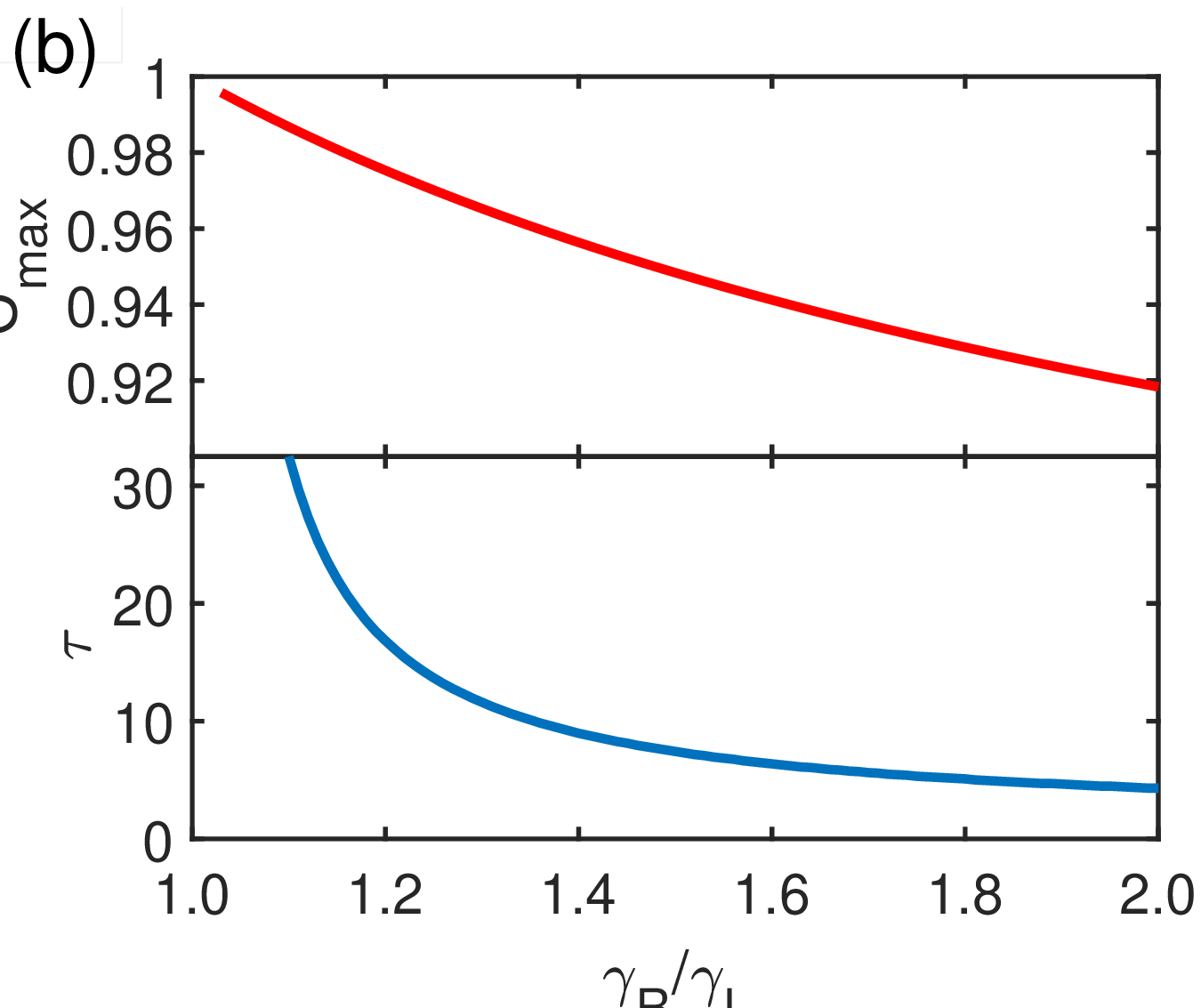}
\includegraphics[width=5.6cm]{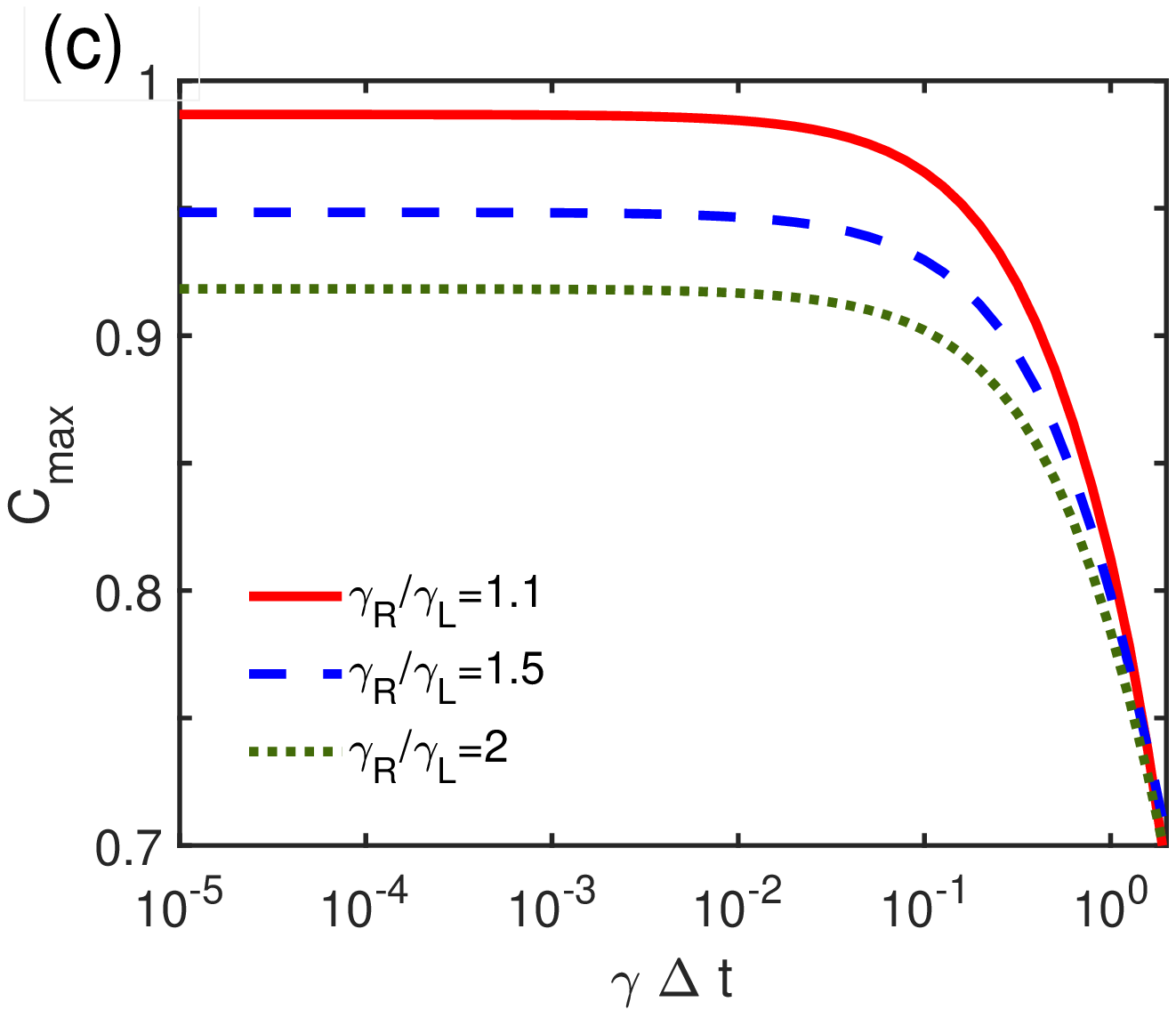}

\caption{\label{chiral2}
 Spontaneous generation of highly entangled states when the retardation effect is negligible. (a) Concurrence as a function of time $\gamma t$
 and chirality $\gamma_R/\gamma_L$. (b) Concurrence (top) and occurrence time (bottom) of maximal achievable entanglement as a function of chirality
 $\gamma_R/\gamma_L$. (c) Maximum concurrence as a function of time delay $\gamma \Delta t$ of a photon propagating from one qubit to the other.}
 \end{figure*}

Furthermore, the coherent feedback can preserve the generated entanglement for appropriate values of chirality $\gamma_R/\gamma_L$.
Fig.~\ref{chiral1}(b) shows how entanglement is preserved as a function of time for different extent of chirality, where we set the
time delay $t_d\approx\tau$ and the qubit-qubit separation $d=\lambda_0/8$. In this case, due to the partial destructive interference between
the emitted and feedback photons, both the qubit-field coupling and the qubit-qubit coupling are weakened. For non-chiral coupling case, qubit 1 is bounded \cite{tufarelli,kim}, such that the qubit-qubit interaction vanishes and the disentanglement process is slightly accelerated. On the contrary, in chiral coupling cases the feedback can preserve entanglement at time longer than its natural disappearance time. For instance, when the coupling is almost unidirectional ($\gamma_R/\gamma_L=20$), entanglement disappears at time $t_{\mathrm{death}}\approx 19/\gamma$ for the semi-infinite waveguide setup, while for the infinite waveguide setup $t_{\mathrm{death}}\approx 7/\gamma$. Moreover, for the intermediate values of chirality ($0<\gamma_R/\gamma_L<2.5$), we have $\mathrm{det}(\textbf{A})\approx0$, where $\textbf{A}$ is a $2\times2$ matrix constructed by the coefficients of $\alpha$ and $\beta$ in Eqs.~(\ref{7}) and (\ref{8}), i.e.
\begin{eqnarray}
\textbf{A}_{11}&=&-\frac{\gamma_L+\gamma_R}{2}+\sqrt{\gamma_L\gamma_R}\;e^{ik_02x_1}, \nonumber\\
\textbf{A}_{12}&=&-\gamma_Le^{ik_0(x_2-x_1)}+\sqrt{\gamma_L\gamma_R}\;e^{ik_0(x_1+x_2)},\nonumber\\
\textbf{A}_{21}&=&-\gamma_Re^{ik_0(x_2-x_1)}+\sqrt{\gamma_L\gamma_R}\;e^{ik_0(x_1+x_2)},\nonumber\\
\textbf{A}_{22}&=&-\frac{\gamma_L+\gamma_R}{2}\beta(t)+\sqrt{\gamma_L\gamma_R}\;e^{ik_02x_2}.
\end{eqnarray}
Consequently, the probability amplitudes $\alpha$
and $\beta$ would evolve autonomously to a stable equilibrium point $(\alpha_e,\beta_e)$, where $\alpha_e/\beta_e\approx-\textbf{A}_{12}/\textbf{A}_{11}$ \cite{breda}. As the determinant of $\textbf{A}$ is not exactly zero, $\alpha$
and $\beta$ still decay at a rate that is much smaller than $\gamma$, thereby a quasi-steady entangled state is realized. Physically, at the equilibrium point, the photonic wavepackets emitted by the two qubits and their reflected counterparts interfere destructively both at positions $x=x_1$ and $x=x_2$, leaving the qubits almost bounded simultaneously. As shown in Fig.~\ref{chiral1}(b), when we set $\gamma_R/\gamma_L=2$, the feedback keeps the concurrence $C>0.2$ at time $t>20/\gamma$, much longer than that in the mirrorless case where $t_{\mathrm{death}}\approx 10/\gamma$.

On the other hand, when the time delay becomes negligible, the system enters from non-Markovian regime to Markovian regime. Remarkably,
in the limit that $t_d\rightarrow0$, or equivalently, qubit 1 is placed almost at the finite end of the waveguide, the coherent feedback
can significantly enhance the maximal achievable entanglement when $\gamma_L$ is sufficiently close (but not equal) to $\gamma_R$.
In this case, the retardation effect is negligible, so that Eqs.~(\ref{7}) and (\ref{8}) can be written as
\begin{eqnarray}
\label{10}
\dot{\alpha}(t)\approx&&-\frac{(\sqrt{\gamma_R}-\sqrt{\gamma_L})^2}{2}\alpha(t)+\sqrt{\gamma_L}(\sqrt{\gamma_R}-\sqrt{\gamma_L})\beta(t),\nonumber\\
\\ \label{11}
\dot{\beta}(t)\approx&&-\frac{(\sqrt{\gamma_R}-\sqrt{\gamma_L})^2}{2}\beta(t)-\sqrt{\gamma_R}(\sqrt{\gamma_R}-\sqrt{\gamma_L})\alpha(t),\nonumber\\
\end{eqnarray}
where we have assumed that the qubit-qubit separation $\Delta x\equiv x_2-x_1=\lambda_0$. Thanks to the coherent feedback, which once again induce a destructive interference between the transmitted and reflected photons, the effective qubit-qubit
coupling strength $g_{\mathrm{eff}}\geq\min(\gamma_L,\gamma_R)\times|\sqrt{\gamma_R}-\sqrt{\gamma_L}|$ , and the effective decay rate
$\gamma_{\mathrm{eff}}=(\sqrt{\gamma_R}-\sqrt{\gamma_L})^2$. When we set $\gamma_L$ to be arbitrarily close to $\gamma_R$, $\gamma_{\mathrm{eff}}$ decreases more rapidly than $g_{\mathrm{eff}}$. It then follows that the ratio
$g_{\mathrm{eff}}/\gamma_{\mathrm{eff}}\approx \sqrt{\gamma_L}/|\sqrt{\gamma_R}-\sqrt{\gamma_L}| \gg1$, which means that the two-qubit
reduced system evolves within the strong-coupling regime. This result is in contrast to previous studies
\cite{vidal,cano,gonzalez,ballestero,facchi,moreno}, in which  $g_{\mathrm{eff}}$ and  $\gamma_{\mathrm{eff}}$ are of the same order
(intermediate-coupling regime). Therefore, as $\gamma_L\rightarrow\gamma_R$, the qubits can readily exchange quanta and get highly
entangled, with the dissipation effect negligible. Interestingly, in the limit that $\gamma_L=\gamma_R$, the qubits are bounded due to disappearance of $g_{\mathrm{eff}}$ and $\gamma_{\mathrm{eff}}$, so that no entanglement can be generated in this case.

In Fig.~\ref{chiral2} (a), we set $t_d=0$ and plot concurrence as a function of chirality $\gamma_R/\gamma_L$ and time $\gamma t$.
As expected, the qubits can get highly entangled when the time delay is negligible. The maximum concurrence
$C_\mathrm{max}$ can be enhanced as $\gamma_R/\gamma_L\rightarrow1$, at the expense that the effective coupling strength $g_\mathrm{eff}$
decreases accordingly and longer time is needed for the emergence of $C_\mathrm{max}$. Fig.~\ref {chiral2} (b) shows maximum concurrence and
its occurrence time $\tau$ as a function of chirality $\gamma_R/\gamma_L$. Clearly, $C_\mathrm{max}$ increases monotonically as $\gamma_R/\gamma_L\rightarrow1$. For
$\gamma_R/\gamma_L=1.1$, $C_\mathrm{max}\approx0.99$ at time $\tau\approx33/\gamma$, that is,
the qubits are almost maximally entangled. Even if $\gamma_R/\gamma_L=2$, $C_\mathrm{max}$ still reaches to 0.92 within a considerably
shorter time with $\tau\approx 4.3/\gamma$. Compared to previous study in Ref. \cite{moreno}, where the maximum concurrence obtained
in infinite waveguide is 0.73, the coherent feedback has significantly enhance the maximal achievable entanglement.

Notice the periodicity of the phases in Eqs.~(\ref{7}) and (\ref{8}), the above results remain valid when $\Delta x$ is an integer multiple of $\lambda_0$, as long as the time delay $\Delta t \equiv \Delta x/v$ that a photon travels from one qubit to the other is negligible. Fig.~\ref{chiral2} (c) shows the dependence of maximum concurrence on the time delay $\gamma\Delta t$, with qubit 2 always placed at the node
of the central EM mode. Similar to the mirrorless case \cite{moreno}, while maximum concurrence decreases as the distance beween the qubits increases, $C_\mathrm{max}$ remains above $0.9$ when $\gamma\Delta t\approx0.01$. For chiral photonic crystal waveguide systems \cite{song}, this
implies that the qubits can be highly entangled even if they are far separated with the distance $\Delta x> 1 \mathrm{cm}$.

\section{\label{sec:level4}EXPERIMENTAL DISCUSSIONS}

In the above section, we have shown that the qubits can get highly entangled spontaneously (with maximum concurrence $C_{\mathrm{max}}>0.9$)
under the conditions that: (i). qubit 1 is placed at the finite end of the waveguide, i.e. the time delay of the feedback $t_d=0$;
(ii). the qubit-qubit separation $d=\lambda_0$; (iii). the decay rates $\gamma_L$ and $\gamma_R$ are close enough with $\gamma_R/\gamma_L\leq2$.
In this section, we adopt the experimental parameters reported in Ref. \cite{song}
and give a brief discussion on how the variation of different parameters affects the generated entanglement.

\begin{figure}[t]

\includegraphics[width=8.1cm]{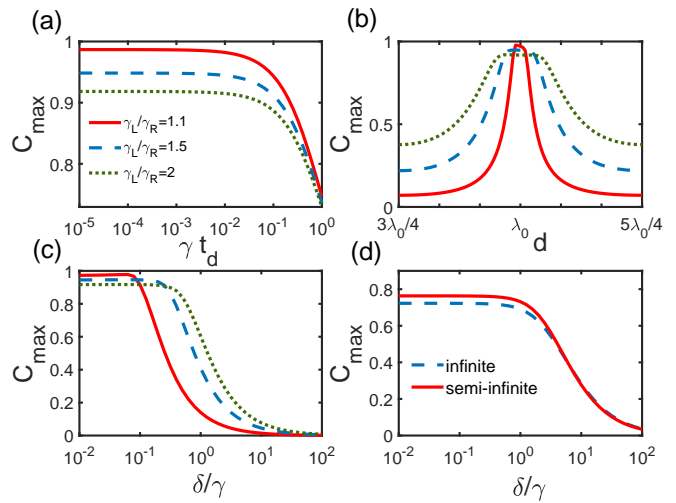}
\caption{\label{chiral5} Maximum concurrence $C_\mathrm{max}$ for different values of chirality $\gamma_R/\gamma_L$. (a) As a function
of time delay $\gamma t_d$. (b) As a function of qubit-qubit separation $d$. (c) As a function of detuning $\delta/\gamma$. (d) Maximum
concurrence $C_\mathrm{max}$ in semi-infinite waveguide (with $\gamma_R/\gamma_L\rightarrow\infty$) and infinite waveguide
(with $\gamma_R/\gamma_L=19$).}
\end{figure}

In Fig.~\ref{chiral5}(a), we plot maximum concurrence $C_\mathrm{max}$ as a function of time delay $\gamma t_d$. For low values of time delay,
$C_\mathrm{max}$ remains unchanged. As the time delay increases such that $\gamma t_d>10^{-2}$, $C_\mathrm{max}$ starts to decrease significantly.
Nevertheless, while lower values of chirality $\gamma_R/\gamma_L$ induce higher values of concurrence, the robustness of
$C_\mathrm{max}$ against time delay $\gamma t_d$ increases as $\gamma_R/\gamma_L$ increases. This effect can also be observed in
Fig.~\ref{chiral5}(b), where the variation of $C_\mathrm{max}$ with respect to the qubit-qubit separation is displayed. We notice that
maximum concurrence reaches its peak value when the qubit-qubit separation $d\approx\lambda_0$, i.e., qubit 2 is also placed at
the node of the central EM mode. Again, higher extent of chirality leads to higher robustness against deviation of qubit 2 from
the optimal position. For example, in order that $C_\mathrm{max}\geq0.9$, one need impose the requirement that the deviation
$\Delta d\leq \lambda_0/20 $ for $\gamma_R/\gamma_L=2$, whereas the upper bound reduces to approximately $\lambda_0/28$ for
$\gamma_R/\gamma_L=1.5$, and $\lambda_0/66$ for $\gamma_R/\gamma_L=1.5$.

Fig.~\ref{chiral5}(c) shows the influence of detuning between the qubits on the maximum concurrence. Following Ref. \cite{moreno},
here we assume that the transition frequencies are modified according to $\omega_1=\omega_0+\delta/2$ and
$\omega_2=\omega_0-\delta/2$ for qubit 1 and qubit 2, respectively. Numerical results suggest that concurrence decreases
monotonically as detuning increases, and, similar to the infinite waveguide case \cite{moreno},
$C_\mathrm{max}$ is more sensitive to detuning when the coupling tends to be non-chiral. Indeed, in the limit that the coupling
tends to be unidirectional to the right (with $\gamma_R/\gamma_L\rightarrow\infty$), the robustness of $C_\mathrm{max}$ is roughly
the same as that in the infinite waveguide case reported in Ref. \cite{moreno} (with $\gamma_R/\gamma_L=19$),
as illustrated in Fig.~\ref{chiral5}(d). However, due to the coherent feedback, maximum concurrence in our model is still slightly
above that in the mirrorless case for most detunings.

\section{\label{sec:level4} CONCLUSION}

In conclusion, we have studied the spontaneous generation of two-qubit entanglement in a semi-infinite waveguide setup.
A key feature of our model, compared to the infinite waveguide case, is the introduction of coherent quantum feedback, which can
be applied to enhance either the robustness or the extent of the generated entanglement. We show that, apart from the non-Markovian
entanglement revival phenomena, the coherent feedback can protect entanglement such that the entanglement death time is significantly delayed.
Moreover, when the retardation effect is negligible and the qubit-waveguide coupling tends to be non-chiral, the two-qubit reduced
system evolves
within the strong-coupling regime and the qubits can be almost maximally entangled. We also study the robustness of our protocol
against variation of some relevant parameters. It is shown that, while the maximum concurrence increases as the
the extent of chirality decreases, the robustness of entanglement decreases accordingly. Thus, experimentally one need
to find a tradeoff between maximal achievable entanglement and robustness of the protocol.

\begin{acknowledgments}
This work is supported by the the Research Foundation of Education Bureau of Fujian Province, China (Grant No. JAT170394,
No. JAT170395, No. JAT170395), Ph.D. research startup foundation of Fujian University of Technology (No. GY-Z13108).
\end{acknowledgments}

\appendix

\section{DERIVATION OF Eqs.~(\ref{7}) AND (\ref{8})}
 We make the transformation \cite{tufarelli} that $\alpha(t)\rightarrow\alpha(t)e^{-i\omega_0t}$, $\beta(t)\rightarrow\beta(t)e^{-i\omega_0t}$, and $\phi(k,t)\rightarrow\phi(k,t)e^{-i\omega_0t}$. Then the first terms in Eqs.~(\ref{4}) and (\ref{5}) vanish, and Eq.~(\ref{6}) can be integrated as
\begin{eqnarray}
\phi(k,t)=&&-i\int_0^t \mathrm{d}s\times e^{i(\omega_k-\omega_0)(s-t)} \bigg(\big(G^*_{ k\scriptscriptstyle L1}+G^*_{ k\scriptscriptstyle R1}\big)\alpha(s) \nonumber\\
&&+ \big(G^*_{ k\scriptscriptstyle L2}+G^*_{ k\scriptscriptstyle R2} \big)\beta(s)\bigg).
\end{eqnarray}
By inserting this integral form into Eq.~(\ref{4}), we find
\begin{widetext}
\begin{eqnarray}
\dot{\alpha}(t)=&&-\int_0^t \mathrm{d}s\; e^{-ik_0v(s-t)}\times \int_{-\infty}^{+\infty}\mathrm{d}k \;e^{ikv(s-t)}\bigg(\big(G_{ k\scriptscriptstyle L1}G^*_{ k\scriptscriptstyle L1}+G_{ k\scriptscriptstyle L1}G^*_{ k\scriptscriptstyle R1}+G_{ k\scriptscriptstyle R1}G^*_{ k\scriptscriptstyle L1}+G_{ k\scriptscriptstyle R1}G^*_{ k\scriptscriptstyle R1} \big)\alpha(s)\nonumber\\
&&+ \big(G_{ k\scriptscriptstyle L1}G^*_{ k\scriptscriptstyle L2}+G_{ k\scriptscriptstyle L1}G^*_{ k\scriptscriptstyle R2}+G_{ k\scriptscriptstyle R1}G^*_{ k\scriptscriptstyle L2}+G_{ k\scriptscriptstyle R1}G^*_{ k\scriptscriptstyle R2} \big)\beta(s)\bigg)\nonumber\\
=&&-\int_0^t \mathrm{d}s\; e^{-ik_0v(s-t)}\times \frac{1}{2\pi}\int_{-\infty}^{+\infty}\mathrm{d}(kv) \bigg(\big(\gamma_Le^{ikv(s-t)}-\sqrt{\gamma_L\gamma_R}e^{ikv(s-t+\frac{2x_1}{v})}-\sqrt{\gamma_L\gamma_R}e^{ikv(s-t-\frac{2x_1}{v})}\nonumber\\
&&+\gamma_Re^{ikv(s-t)}\big)\alpha(s)+
\big(\gamma_Le^{ikv(s-t-\frac{x_1-x_2}{v})}-\sqrt{\gamma_L\gamma_R}e^{ikv(s-t-\frac{x_1+x_2}{v})}-
\sqrt{\gamma_L\gamma_R}e^{ikv(s-t+\frac{x_1+x_2}{v})}\nonumber\\
&&+\gamma_Re^{ikv(s-t+\frac{x_1-x_2}{v})}\big)\beta(s)\bigg)\nonumber\\
=&&-\int_0^t \mathrm{d}s\; e^{-ik_0v(s-t)}\bigg(\big(\gamma_L\delta(s-t)-\sqrt{\gamma_L\gamma_R}\;\delta(s-t+\frac{2x_1}{v})
-\sqrt{\gamma_L\gamma_R}\;\delta(s-t-\frac{2x_1}{v})+\gamma_R\delta(s-t)\big)\alpha(s)\nonumber\\
&&+\big(\gamma_L\delta(s-t-\frac{x_1-x_2}{v})
-\sqrt{\gamma_L\gamma_R}\;\delta(s-t+\frac{x_1+x_2}{v})
-\sqrt{\gamma_L\gamma_R}\;\delta(s-t-\frac{x_1+x_2}{v})\nonumber\\
&&+\gamma_R\delta(s-t+\frac{x_1-x_2}{v})\big)\beta(s)\bigg)\nonumber\\
=&&-\frac{\gamma_L+\gamma_R}{2}\alpha(t)-\gamma_Le^{ik_0(x_2-x_1)}\beta(t-\frac{x_2-x_1}{v})\theta(t-\frac{x_2-x_1}{v})+
\sqrt{\gamma_L\gamma_R}\;e^{ik_02x_1}\alpha(t-2x_1/v) \nonumber\\
&&\times\theta(t-2x_1/v)+\sqrt{\gamma_L\gamma_R}\;e^{ik_0(x_1+x_2)}\beta(t-\frac{x_1+x_2}{v})\theta(t-\frac{x_1+x_2}{v}),
\end{eqnarray}
\end{widetext}
where we have used the fact that $\omega_k=kv$. This completes the derivation of Eq.~(\ref{7}), and one can derive Equation (\ref{8}) in a similar way.

\end{document}